\documentclass[useAMS,usenatbib]{mn2e_astroph}

\usepackage{graphicx}

\def\nct#1{\nocite{#1}}
\newcommand{\rlc}{R_{\rm lc}}
\def \rat{R_\rho}
\def \mref#1{(\ref{#1})}

\title[Peak separation ratio as a probe of pulsar beam]
{The ratio of profile peak separations 
as a probe of pulsar radio-beam structure}

\author[J.~Dyks and M.~Pierbattista]{J.~Dyks and M.~Pierbattista
\\
Nicolaus Copernicus Astronomical Center, Rabia\'nska 8, 87-100, Toru\'n,
Poland}
\begin{document}

\date{Accepted .... Received 2015 May 25; in original form 2015 May 23}


\maketitle

\label{firstpage}

\begin{abstract}
The known population of pulsars contains objects with four 
and five component profiles, for which the peak-to-peak separations 
between the inner and outer components can be measured.
These Q and M type profiles can be interpreted as
a result of sightline cut through a nested cone beam, or through a set of
azimuthal fan beams.
We show that the ratio $R_W$ of the components' separations provides 
a useful measure 
of the beam shape, which is mostly independent of
parameters that determine the beam scale and
 complicate interpretation of simpler profiles.
 In particular, the method does not depend on 
the emission altitude and the dipole tilt distribution. 
The different structures of the radio beam imply manifestly
different statistical distributions of $R_W$, with the conal model 
being several orders of magnitude less consistent with data than the
fan beam model. 
To bring the conal model into consistency with data,
strong effects of observational selection need to be called for,
with 80\% of Q and M profiles assumed to be undetected because 
of intrinsic blending effects.
It is concluded that the statistical properties of Q and M profiles
are more consistent with the fan-shaped beams,
than with the traditional nested cone geometry.
\end{abstract}


\def\lap{\hbox{\hspace{4.3mm}}
         \raise1.5pt \vbox{\moveleft9pt\hbox{$<$}}
         \lower1.5pt \vbox{\moveleft9pt\hbox{$\sim$ }}
         \hbox{\hskip 0.02mm}}

\def\rwobs{R_W}
\def\rwcon{R_W}
\def\rwstr{R_W}
\def\winobs{W_{\rm in}}
\def\woutobs{W_{\rm out}}
\def\phm{\phi_m}
\def\phmi{\phi_{m, i}}
\def\thm{\theta_m}
\def\dres{\Delta\phi_{\rm res}}
\def\win{W_{\rm in}}
\def\wout{W_{\rm out}}
\def\rin{\rho_{\rm in}}
\def\rout{\rho_{\rm out}}
\def\phin{\phi_{\rm in}}
\def\phout{\phi_{\rm out}}
\def\xin{x_{\rm in}}
\def\xout{x_{\rm out}}

\def\thmin{\theta_{\rm min}^{\thinspace m}}
\def\thmax{\theta_{\rm max}^{\thinspace m}}

\section{Introduction}
\label{intro}

In spite of a large and increasing number of detected radio pulse profiles,
a generic shape of pulsar beam remains a subject of debate. 
The mainstream 
models seem to support patchy or conal geometry. The patchy 
form is supported by the diversity and asymmetry of profiles, as well 
as by the invoked distribution of individual components within the polar 
tube (Lyne \& Manchester 1988; Manchester 2012).
\nocite{lm88,man12}

In a series of papers, (Rankin 1988; 1990; 1993, hereafter R93), 
\nocite{ran83}\nocite{ran90}\nocite{ran93}
Joanna Rankin provides arguments for approximate beam geometry in 
the form of two nested cones with an axial core component. It has been 
suggested that this beam geometry is approximately universal, with many 
pulsars having either the inner or outer cone, with a possible core component.
A small group of profiles with 4 and 5 components (Q and M type,
respectively) has been interpreted as a cut of sightline through both cones. 
The central component in M-type profiles is created by the additional 
co-axial core beam.
In the works of J.~Rankin,
the angular radii of the cones
have been estimated for pulse longitudes measured at the outer $50$\%
peak flux of components. At 1 GHz they are equal to: 
$\rho_\mathrm{in}^{50}=4.3^\circ P^{-1/2}$ 
and $\rho_\mathrm{out}^{50}=5.8^\circ P^{-1/2}$ 
for the inner and outer cone, respectively; ($P$ denotes
a pulsar period, which in these equations should be specified in seconds).
The result has been confirmed by other groups, who measured
the conal pair widths at a different flux level and frequency 
(Gil et al.~1993, hereafter G93; Kramer et al.~1994, hereafter K94;
Mitra \& Deshpande 1999, hereafter MD).
\nocite{gks93} 
\nocite{kwj94}
\nocite{md99}
MD have tentatively identified three cones, 
out of which we select the inner two,
because they outnumber the third-cone case, 
and their ratio is consistent
with that derived in other studies.

\cite{wri03} 
introduced a special-relativistic model of drifting 
pulsar beams in which the two cones are associated with two particular sets of dipolar 
magnetic field lines: the last open lines and the critical lines, which
separate zones of opposite-sign charge at the light cylinder
 (of radius $\rlc=cP/(2\pi)$, where $c$ is the vacuum speed of light). 
In pulsar magnetospheres,
these lines form two co-axial tube-shaped surfaces
with different opening angles, $2\rho_{\rm lo}$ and $2\rho_{\rm crt}$.
In a dipolar field with magnetic moment parallel to the rotation axis,
it holds that $\rho_{\rm cr}(r)/\rho_{\rm lo}(r) = 0.75$. 
If the radial distance $r$ 
from the center of a neutron star is not too large ($r\ll\rlc$), the result is
independent of $r$.
For a large tilt of static-shaped dipole, the theoretical ratio increases
up to $0.82$.
\cite{wri03} notes that
the numbers are close to the size ratio of the cones observed by 
Rankin (1993):
\nct{ran93} 
$R_\rho^{50} = \rho_\mathrm{in}^{50}/\rho_\mathrm{out}^{50} = 0.74$.
In Table \ref{tab1} we give other values of the ratio, 
 as determined from observations by several research groups. 
\begin{table*}
\begin{tabular}{|c|ccc|ccc|ccc|}
\hline
 & MD & R93 & G93 & \multicolumn{3}{c|}{K94 ($P^{-\kappa}$)} & 
               \multicolumn{3}{c|}{K94 ($P^{-0.5}$)} \\
\hline
$\nu$ [GHz] & 1.0 &  1.0 & 1.4 & 1.4 & 4.75 & 10.55 & 1.4 & 4.75 & 10.55 \\
$R_F$ [\%] & 100 &  50 & 10 & 10 & 10 & 10 & 10 & 10 & 10 \\
\hline
$\rho_\mathrm{in}$ [$^\circ$] & 4.1 & 4.3 & 4.9 & 5.3 & 4.5 & 4.77 & 4.9 & 4.4 & 4.5 \\
$\rho_\mathrm{out}$ [$^\circ$] & 5.1 & 5.8 & 6.3 & 6.23 & 5.76 & 5.48 & 6.3 & 5.9 & 5.5 \\
$\rat$ & 0.8 & 0.74 & 0.78 & 0.85 & 0.78 & 0.87 & 0.78 & 0.75 & 0.82 \\
\hline
\end{tabular}
\caption{Angular radius of the inner ($\rho_\mathrm{in}$) 
and outer ($\rho_\mathrm{out}$) cone, and 
their ratio $\rat=\rho_\mathrm{in}/\rho_\mathrm{out}$, 
as determined by various statistical studies of pulsar profiles.
The values are based on profile widths measured
at a different fraction $R_F$ of the components' peak flux.
In the case of MD, who have identified three cones, 
we provide values for the inner two.
The last three columns (from K94) and the values from MD
refer to a fit with a fixed 
period dependence of $P^{-0.5}$. 
}\label{tab1}
\end{table*}
\nocite{ran90,ran93}
\nocite{gks93} 
\nocite{kwj94}
\nocite{md99}
A narrowing of cones with increasing frequency $\nu$,
can be inferred from the last 6 columns of Tab.~\ref{tab1}.
In spite of this, the ratio of cones' size $\rat$
remains $\nu$-independent. This is consistent with the cones
occupying the same magnetic field lines at different altitudes.
The geometry of the emission region seems to follow the flaring geometry
of the dipolar magnetic field.

The cone
size ratio $\rat$
depends on the flux level at which
locations of components in a profile are measured.
This is usually set as a fraction $R_F$ of peak flux of a considered 
component (see Fig.~4 in Kramer et al.~1994).
For similar components of the inner and
outer cone (similar width and shape), $\rat$ should slightly increase
with decreasing $R_F$. Comparison of results from R93, G93 and K94 
\nocite{ran90,ran93}
\nocite{gks93} 
\nocite{kwj94}
suggests a weak increase of $\rat$ for $R_F$ decreasing from 50 to 10\%.
Columns 2-4, however, which also include the result of MD for
$R_F = 100$\%, provide little evidence for this. 
In this paper we deal with the 
multicomponent profiles of class Q and M in which components often
partially overlap with each other.
We find that peaks of such blended components can, on average, be more easily
identified than the points corresponding to a lower flux fraction.
Therefore, to minimise
the blending problems, we assume $R_F = 100$\%.

In addition to the patchy and conal beams, a variety of more complicated
shapes have been considered, such as the hourglass shape 
(Weisberg \& Taylor 2002),
\nocite{wt02}
elliptic (MD; Perera et al.~2010),
\nocite{md99}\nocite{pmk10}
 various systems of fan beams, 
eg.~spoke-like or wedge-like 
(Dyks et al.~2010, hereafter DRD10; Wang et al.~2014; 
Teixeira et al.~2016)
\nct{drd10, wpz14, trw16}
and in the form of a modelled polar cap rim
as determined by the magnetic fieldline tangency condition 
at the light cylinder (Biggs 1990; Dyks \& Harding 2004).
\nct{big90,dh04} 
A hybrid form consisting of `patchy cones' has also been considered by
Karastergiou \& Johnston (2007) 
\nct{kj07}
and shown to reproduce some 
statistical properties of pulsar profile ensamble.

There is a growing evidence that pulsar radio beams generally do not have a
conal geometry.
\nct{drd10} 
DRD10
proposed a radio emission beam in the form of multiple fan
beams created by plasma streams diverging from the magnetic dipole axis.
This radio emission geometry,
 resembling the pattern of spokes in a wheel when viewed down the dipole
axis, has shown many advantages when compared to the 
nested-cone case (see Dyks and Rudak 2012; Desvignes et al.~2013; 
\nct{dr12, dr15, dkc13, wpz14}
Wang et al.~2014; Dyks \& Rudak 
2015).
The new model has managed to explain the main frequency-dependent
features of multicomponent profiles, such as the radius-to-frequency mapping
and the relativistic core lag (Gangadhara \& Gupta 2001). \nct{gg01}
It also provides a successful interpretation
of double notches observed in some averaged profiles (McLaughlin \& Rankin
2004). \nct{mr04}

With the long-term monitoring, 
pulsar beams can be mapped for precessing objects,
especially those which undergone the fast precession caused by the
relativistic spin-orbit coupling (Kramer 1998; Hotan et al.~2005;
Clifton \& Weisberg 2008). 
\nct{kra98,hbo05,cw08} 
The rare examples that have been mapped so far show that
there is much to learn about the form of pulsar beams.
Beam maps in Desvignes et al.~(2013) and Manchester et al.~(2010)
\nct{dkc13,mks10}
suggest elongated patterns that point at the magnetic pole, in line
with the fan-shaped
pattern discussed in DRD10 (fig.~18 therein). \nct{drd10}

The majority of previous works 
have focused on a study of widths of radio 
profiles.
The widths, however, are sensitive to a number of factors, 
such as the rotation period $P$, physical size of the emission region,
altitude of emission, and the macroscopic pulsar geometry 
(dipole tilt and viewing
angle). All of them are convolved and, except from the period, unknown 
for the majority of objects. 

The Q and M profiles provide a useful tool for deciphering the 
pulsar beam shape, because it is possible to study the \emph{ratio}
of separations between their components, instead of the full width of a
profile.
With a good accuracy, distributions of such ratios are insensitive to
major parameters that complicate the analysis of profile widths
(such as the emission altitude, frequency $\nu$, 
rotation period $P$, and the specific form of a dipole tilt 
distribution).
In this paper we study the statistics of components' separation ratio
for the conal and fan-beam models, and compare with the observed
distribution.
 
The outline of this paper is the following: 
In Section \ref{conmod} we describe the main idea of the paper,
and apply it to the conal model.
Section \ref{observed} presents the way in which the
observed distribution of the peak-separation ratio was
 determined. In Section \ref{concom} we compare
the theoretical
distribution of the peak-separation ratio in the nested cone model 
with the observations. 
Section
\ref{Peak separation stream} does the same for the stream-like 
geometry, and is followed by discussion and conclusions.

\section{Peak-separation ratio in the nested cone model}
\label{conmod}

\begin{figure}
\includegraphics[width=0.48\textwidth]{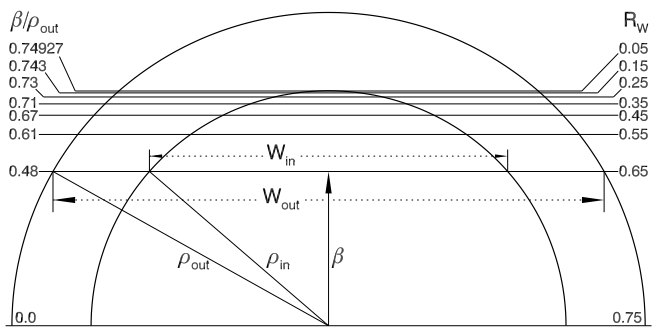}
\caption	{Top half of a nested cone beam (two half-circles) 
with the cone size ratio 
$R_\rho = 0.75$. The set of horizontal lines (paths of different sightlines) 
presents intervals of viewing angle
that correspond to a fixed interval 
of the peak-separation ratio $\Delta R_W = 0.1$.
Values of the impact angle $\beta$ (in units of the outer conal radius
$\rout$) are given on the left, values of $R_W$ -- on the right hand side.
Note that it is a lot more probable to observe $R_W\in(0.65, 0.75)$
than the smaller values. 
}
\label{idea}
\end{figure}

 The idea is based on the measurement of
 the peak-to-peak width
$\win$ and $\wout$, for the inner
and outer pair of conal components, respectively, 
to derive the width ratio $R_W = \win/\wout$.
For a central cut through the nested cones 
with angular radii
$\rin$ and $\rout$,
the ratio $R_W$ is maximal
and equal to the ratio of cones' size: $\rat=\rin/\rout$.
This is also the most likely value, which should
be vastly dominant in the data. Fig.~\ref{idea} presents the upper 
half of a nested
cone beam with $\rat=0.75$, viewed down the dipole axis.
The horizontal lines mark the paths of the line of sight for observers
located at different impact angles $\beta = \zeta - \alpha$, where
$\zeta$ is the angle between the sightline and rotation axis $\vec \Omega$,
and $\alpha$ is the tilt of a magnetic dipole with respect to $\vec \Omega$.
The real spherical geometry of the sightline cut implies that
the viewing paths are curved in general.
However, the straight lines of Fig.~\ref{idea} provide 
a good approximation
whenever $\alpha \gg \rout$ and $\rout \ll 1$ rad.
The paths have been plotted for the \emph{equidistant} values of the peak
separation ratio $R_W$, printed on the right-hand side. Adjacent values of
of $\beta/\rout$ (shown on the left) quickly approach each other
with decreasing $R_W$, which means that the chance to observe
the inner components at a small separation is considerably smaller
than observing them at a larger distance.
For example, the probability to observe $R_W$ in the range $(0.25, 0.35)$
is $(0.73-0.71)/0.48=0.04$ times smaller than to observe a larger $R_W$
in the same-width interval of $(0.65,0.75)$.

For a nested cone beam with a universal $\rat$, 
the distribution of $R_W$ is a single-peaked function, monotonously increasing
towards the sharp peak at $R_W = \rat$. 
As demonstrated in Appendix \ref{flatderiv},
the function $n(R_W)$
can be easily derived in the flat case shown in Fig.~\ref{idea}: 
\begin{equation}
n(R_W) = C\left|\frac{d n}{d R_W}\right| = C \rout
\frac{R_W(1- \rat^2)}{(\rat^2-R_W^2)^{1/2}(1-R_W^2)^{3/2}}
\label{nodrw}
\end{equation}
where $C$ is a normalisation constant. 

This distribution has important
advantages 
over a direct statistical study of
pulse widths, because $\win$ and $\wout$ are expected to depend on 
emission altitude, observation frequency, rotation period of the star,
and the dipole tilt $\alpha$.
In the case of small beams, typical of ordinary pulsars,
the flat case of eq.~\mref{nodrw} presents a good approximation
within
most of the $(\alpha, \zeta)$ parameter space, except from cases of 
the nearly aligned geometry. The latter, however, are not numerous,
so the actual distribution $n(R_W)$, as calculated with 
the strict account of the spherical trigonometry, is close to
eq.~\mref{nodrw}.
Therefore, it is worth to discuss the properties of the 
$R_W$ distribution
within the range of validity of eq.~\mref{nodrw}.
If the cones are associated with the last open and critical field lines
(or any lines defined by two fixed values of the footprint 
parameter\footnote{The footprint parameter is a ratio of the magnetic 
colatitude of an arbitrary point and the magnetic colatitude 
of the open field line boundary, measured at that point's radial distance.}),
then the value of $R_W$ is not expected to depend 
on the rotation period P or the emission altitude $r$.
This is because a change in $P$ or $r$
just rescales the beam shown in Fig.~\ref{idea}, 
with no influence on the relative geometry 
of cones and the statistics of $R_W$. For the aforementioned
$B$-field lines, the ratio $\rin/\rout$ 
does not depend on $r$ as long as $r\ll \rlc$. Therefore, we fix it at
$R_\rho=0.75$.
A choice of frequency $\nu$ should not affect $R_W$ either, 
if the peak emission at different $\nu$
corresponds to different altitudes, but the same type of dipolar field lines
(last open/critical).
Note that $\rout(P,r)$ in eq.\mref{nodrw} can be incorporated into the
normalisation constant $C$. This is because the fraction in eq.~\mref{nodrw}
(let us denote it by $F$, so that $n=C\rout F$)
is almost insensitive to $P$ and $r$. A `total' distribution 
that incorporates beams of different size $\rho_{\rm out,i}$
can then be written as $n\approx\sum_i C \rho_{\rm out,i} F 
= CF\sum_i\rho_{\rm out,i}=C^\prime F$.
For all the afore-described reasons, the distribution of peak separation
ratio $R_W$ provides a useful, one-dimensional tool for testing
the pulsar beam shapes. It is insensitive (or very weakly sensitive) 
to the uncertain parameters, and allows us to avoid the usual 
two-dimensional analysis (eg.~of $n(W,P)$). On the bad side,
it is applicable only for a limited number of objects with 
four and more components (of Q and M type).

\begin{figure}
\includegraphics[width=0.48\textwidth]{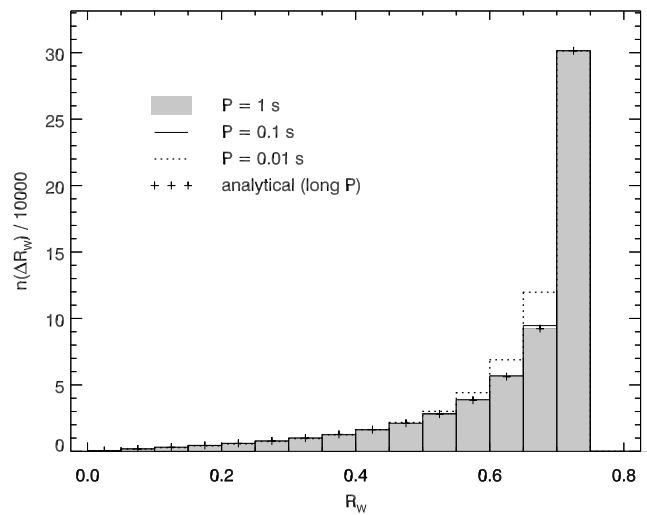}
\caption{Dependence of the $R_W$ distribution on period, 
with $\alpha$ and $\zeta$ sampled isotropically.
Note that the grey histogram ($P = 1$ s) essentially coincides
with the analytical case of eq.~\mref{nodrw} ($P = \infty$, marked with the plus signs), 
despite the former includes
cases with $\alpha \ll 90^\circ$. The solid line case for $P=0.1$ s 
is very close to
the aforementioned (long $P$) cases. All of these distributions are
therefore appropriate for majority of normal pulsars.
The numbers on the vertical axis refer to $P=1$ s, with 
the other histograms normalised at the same peak value. 
}
\label{periods}
\end{figure}

Since Fig.~\ref{idea} and eq.~\mref{nodrw} are only valid 
for large $\alpha\gg \rout$, we have determined the distribution $n(R_W)$
numerically, by calculating exact values of $R_W$ for a large sample
of beams ($10^7$) simulated for isotropically distributed angles of $\alpha$, 
and $\zeta$. 
The opening angle of the outer cone has been calculated with
the usual dipolar formula:
\begin{equation}
\cos\rout = 
(2-3r/\rlc)(4-3r/\rlc)^{-1/2}.
\label{cosrho}
\end{equation}
A variety of periods $P$, typical of normal pulsars has been tried
to verify the near-independence of $n(R_W)$ on $P$ (see Fig.~\ref{periods}).
Strong discrepancy has appeared only for $P$ less than a few tens of
milliseconds (the dotted line in Fig.~\ref{periods} is for $P=10$ ms),
whereas the periods in our sample of observed Q and M pulsars 
range between $0.16$ and $2.25$ s, with an average of $0.84$ s.
For smaller $P$, the pulsar beam is larger, and the curvature of the 
sightline's path within the beam is more important. As shown in Appendix
\ref{alphadep}, this effect is second order in $\rout$, ie.~it is usually small.
Because of the curved viewing paths, $R_W$ decreases, 
hence the $R_W$ values from the
highest histogram bin ($R_W \in \left(0.7,0.75\right)$)
start to pour over to adjacent bins on the left
(with $R_W < 0.7$). See the dotted line in Fig.~\ref{periods}.
This effect is more pronouced for narrower histogram bins.

Following R93, the value of $r$ has been set to $220$ km, and we have 
assumed $\rin = 0.75\rout$. 
The period has been set to $P=1$ s, which is a
round number close to the average $P$ in the observed sample of Q and M 
pulsars. The average $P$ of the total population of known pulsars is smaller, 
however, evidence has been presented for that the Q and M profiles
are mostly observed in old objects (Rankin 1990). 
The resulting pulse width for the inner and outer cone has been
calculated with the spherical cosine theorem 
\begin{equation}
\cos(W_i/2) = (\cos\rho_i - \cos\alpha
\cos\zeta)(\sin\alpha\sin\zeta)^{-1},
\label{cosphi}
\end{equation}
 where the index $i$ refers to `in' or
`out'. Since $\alpha$ and $\zeta$ are blindly sampled 
from an isotropic distribution, for most of them 
the line of sight does not traverse both cones.
This happens only when eq.~\mref{cosphi} gives
 finite solutions for both $W_i$, thus
we accept only those $(\alpha, \zeta)$-pairs for which
\begin{equation}
|\beta| < \rin
\ \ \ \ \ {\rm and} \ \ \ \ \   
\cos{\rout} > \cos{(\alpha + \zeta)}.
\end{equation}
For $P=1$ s and $r=220$ km, $\rin=4.38^\circ$, so the conditions are passed
by just a few percent of total number of cases.

\begin{figure}
\includegraphics[width=0.48\textwidth]{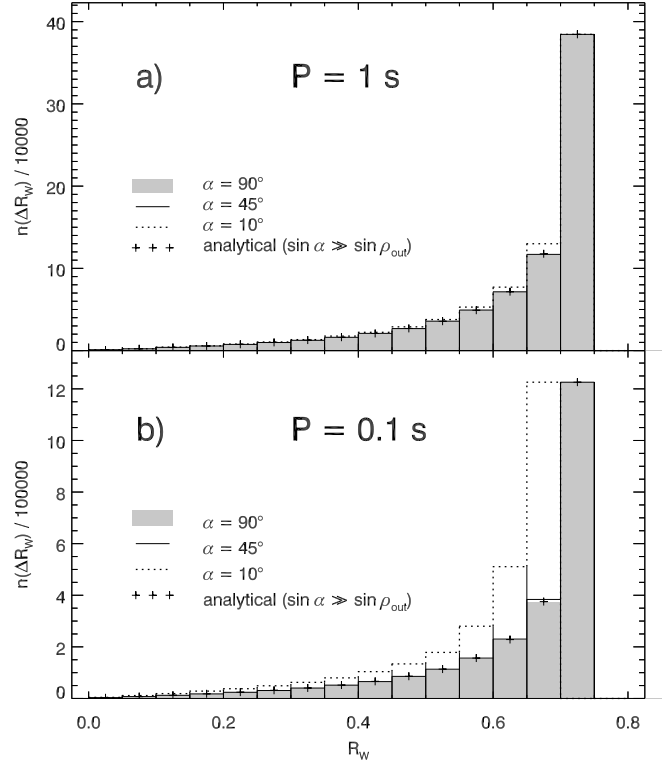}
\caption{Dependence of the $R_W$ distribution with the dipole tilt $\alpha$
for $P=1$ s (top) and $P=0.1$ s (bottom). Only the viewing angle 
$\zeta$ was sampled isotropically.
Note that the $R_W$ distribution practically 
does not depend on $\alpha$, 
except from when 
both
$\alpha=10^\circ$ and $P=0.1$ are simultaneously small (dotted line in
bottom panel) which is a rare circumstance.
}
\label{alphas}
\end{figure}

Fig.~\ref{alphas} presents the $n(R_W)$ distribution
for selected values of dipole inclination $\alpha$
and for two values of $P=1$ s (top) and $0.1$ s (bottom).
In the long-period case (top panel), even moderate dipole inclinations
result in a distribution which is well described by eq.~\mref{nodrw}:
the solid line histogram for $\alpha=45^\circ$ is indiscernible
from the grey orthogonal ($\alpha=90^\circ$), or from the analytical case).
Pronounced difference can only be seen for a very small inclination
$\alpha=10^\circ$ and short $P$ (Fig.~\ref{alphas}b, dotted).
The reason for this is that for decreasing $\alpha$ both $\win$ and $\wout$
scale approximately as $W_i/\sin{\alpha}$ which makes $R_W$ very
stable.
In Appendix \ref{alphadep} a second-order expansion of $R_W$ is made for the special
case of the central cut ($\alpha = \zeta$):  
\begin{equation}
R_W \approx \rat \left( 1 - \frac{1 -
\rat^2}{24}\frac{\rout^2}{\sin^2{\alpha}} \right) = 
\rat \left( 1 - 0.018\frac{\rout^2}{\sin^2{\alpha}} \right),
\label{secorder}
\end{equation}
where the numeric value on the right corresponds to $\rat=0.75$.
 As one can see, $R_W$ starts to perceptibly depend on $\alpha$
only for a nearly-aligned geometry ($\alpha\la\rout$).
According to eq.~\ref{secorder}, when $\alpha$ is decreasing, 
$R_W$ decreases 
with respect to the flat-case value of $\rat=0.75$. Hence
$\win$ is increasing a bit slower than $\wout$.
Because of the square dependence on $\rho_i/\sin\alpha$, a 
considerable divergence may appear only when $\rho_i$ is
relatively\footnote{Eq.~\mref{secorder} has been derived for $\rout \ll 1$
rad and $\sin{\alpha} \gg \rout$.} 
large or $\sin\alpha$ small.
In the rare cases when $\sin\alpha \lap \rout$ 
(dotted line in Fig.~\ref{alphas}b),
a strong discrepancy from the flat analytical case appears
and the histogram does not extend all the way up to $\rat$.
In the case of such a small $\alpha$ the line of sight is capable of staying 
for most of the time between the cones. 
An extreme example with $\win\approx0$ and 
$\wout\approx2\pi$, is the case 
with the $\vec \Omega$ axis located half way between the cones
($\alpha=\rin+(\rout-\rin)/2$) and $\zeta\approx(\rout-\rin)/2$.
However, since such cases are rare they do not affect the total 
$R_W$ distribution which is 
based on the isotropic distribution of $\alpha$.
Therefore, the sensitivity of $n(R_W)$ to the key geometrical 
parameter, the dipole tilt $\alpha$, is marginal, 
and nearly completely reduced
as compared to the sensitivity of the $n(W)$ distribution.

\section{The observed $R_W$-distribution}
\label{observed}

\begin{table*}
\begin{tabular}{l c c c c c c c c c}
Name & Ref. & Type & Qual. & $P$ & Freq. & $\phi_1$, $\phi_2$, $\phi_{n-1}$,
$\phi_n$ & $\win$ & $\wout$ & $R_W$ \\

 & & & & s & GHz & $^\circ$ & $^\circ$ & $^\circ$ & \\
\hline
B0329+54 &   1\nct{hx97} & M & 1 & 0.714 &   1.41  &    33.0, 41.3, 49.8,  54.2  & 8.5 & 21 & 0.40 \\
J0401-7608 & 2\nct{jnk98} & Q & 2 & 0.545 &   0.6  &   -4.2,  -0.7,  3.5,   7.5   & 4.2 & 11.7 & 0.36  \\ 

B0621-04 &   3\nct{gl98} & Q & 1 & 1.039 &  1.408 &  162.5, 168.0, 174.6, 179.8 & 6.6  & 17.3 & 0.38   \\
J0631+1036 & 4, 5\nct{waa10}, 6\nct{trw16} & Q & 1 & 0.287 &   1.4  &     -9.2, -2.8, 3.0, 8.5 & 5.8 & 17.7 & 0.33  \\ 

J0742-2822 & 7\nct{jhv05}, 8\nct{kj07} & M & 2 & 0.166 &   1.5  &    -7.6,  -6.1, -1.5, 2.5 & 4.6 & 10.1 & 0.46 \\ 
J1034-3224 & 4, 9\nct{mhq98} & Q & 2 & 1.150 &   0.6  &   -33.0,  0.0,  14.0, 39.0 & 14.0 & 72 & 0.19\\ 
B1055-52 & 4 & Q & 1 & 0.197 &   3  &   -89.0,  -79.0, -66.3, -59.0 & 12.7 & 30 & 0.42 \\ 

B1237+25 &   10\nct{sgg95} & M & 1 & 1.382 &  1.4   &  42.45, 44.65,  50.7,  52.0 & 6.0  & 9.5 & 0.63  \\
J1326-5859 & 4 & Q & 2 & 0.477 &   3  &       -7.9,  -2.0, -0.2, 5.0  & 1.8 & 12.9 & 0.14 \\ 
J1327-6222 & 8\nct{kj06} & Q & 2 & 0.529 &   3  &  -2.5, 0.0, 3.2, 4.6          & 3.2 & 7.1 & 0.45 \\ 
J1536-3602 & 4 & Q & 2 & 1.319 &   0.6  &  -6.5, 1.0, 7.9, 14.3         & 6.9 & 21 & 0.33\\ 
J1651-7642 & 4 & Q & 2 & 1.755 &   1.5  &    -1.0, 2.5,  6.7, 11.0      & 4.2 & 12.0 & 0.35\\ 

B1737+13 &   3\nct{gl98} & M & 2 & 0.803 &  1.408 &   180.0, 183.5, 192.0, 195.0 & 8.5 & 15 & 0.57 \\
B1738-08 &   3 & Q & 1 & 2.043 &   0.61 &   173.8, 177.0, 182.7, 185.8 & 5.7 & 12 & 0.47\\
B1804-08 &   3 & Q & 1 & 0.163 & 1.408  &  151.0, 158.7, 166.2, 169.3  & 7.5 & 18.3 & 0.41 \\

J1819+1305 & 11\nct{rw08} & Q & 1 & 1.060 &   0.327 &   -11.5, -5.3, 7.2, 11.5      & 12.5 & 23 & 0.54 \\ 
B1821+05   & 12\nct{hr10} & M & 2 & 0.752 &   0.43/1.401  &    -13.0, -6.0, 6.5, 10.5     & 12.5 & 23.5 & 0.53 \\ 

B1831-04 &   3\nct{gl98} & M & 1 & 0.290 &  0.606 &  138.5, 153.0, 212.5, 233.5  & 59  & 95 & 0.63 \\
B1845-01 &   3, 10 & Q & 2 & 0.659 &    1.42 &    62.0,  64.2,  68.4, 74.4 & 4.2 & 12.4 & 0.34 \\
B1857-26 &   3 & M & 2 & 0.612 &  0.61  &  169.0, 176.5, 192.3, 198.5  & 15.8 & 29 & 0.53  \\ 
B1910+20   & 12\nct{hr10} & M & 2 & 2.232 &   1.615 &     -7.0, -3.3, 4.5, 7.4      & 7.8 & 14.4 & 0.54 \\ 

B1918+19   & 13\nct{rwb13} & Q & 2 & 0.821 &   0.327 &    -24.7, -10.7, 1.0, 23.5    & 11.7 & 48 & 0.24 \\ 

J1921+2153 & 4 & Q & 2 & 1.337 &  3  &   -3.7,  -1.8,   1.2,   3.1  & 3.0 & 6.8 & 0.44   \\
B1929+10 &   1, 14\nct{kxj97} & Q & 2 & 0.226 &   1.71 &   205.5, 207.5, 210.5, 213.4 & 3.0 & 7.9 & 0.38 \\
B1952+29 &   3 & Q & 2 & 0.426 &  1.408 &   253.6, 259.0, 264.7, 268.8 & 5.7 & 15.2 & 0.37 \\
B2003-08 &   3 & M & 2 & 0.580 &   0.408 &   162.0, 174.0, 205.0, 214.0 & 31 & 52 & 0.60 \\

B2028+22   & 12 & Q & 1 & 0.630 &   0.430 &      -4.8, -0.7, 2.6, 6.9     & 3.3 & 11.7 & 0.28\\
B2210+29 &   3, 10 & M & 2 & 1.004 &   1.42  &   160.3, 162.6, 171.8, 174.9 & 9.2 & 14.6 & 0.63 \\
B2310+42 &   1 & M & 2 & 0.349 &  1.41 &    106.7, 109.0, 115.3, 117.8 & 6.3 & 11.1 & 0.57\\
B2319+60 &   3 & Q & 2 & 2.256 &   0.925 &   173.9, 179.0, 182.3, 188.8 & 3.3 & 14.9 & 0.22\\ 


\hline
\end{tabular}
\caption{A subset of Q and M-type pulsars for which the peak separation
ratio $R_W$ was possible to estimate. The component phases $\phi_i$ refer to an arbitrary zero phase.
References: 1) von Hoensbroech \& Xilouris (1997);
 2) Johnston et al.~(1998); 3) Gould \& Lyne (1998); 
4) the ATNF database (http://www.atnf.csiro.au/people/joh414/ppdata);
5) Weltevrede et al.~(2010); 6) Teixeira et al.~(2016);
7) Johnston et al.~(2005); 8) Karastergiou \& Johnston (2006);
9) Manchester et al.~(1998); 10) Seiradakis et al.~(1995);
11) Rankin \& Wright (2008); 12) Hankins \& Rankin (2010);
13) Rankin et al.~(2013).
}
\label{tabobs}
\end{table*}

To verify the simulated distribution of $R_W$
we have computed the observed distribution (Fig.~\ref{com}c), 
by estimating the peak separations
for 30 pulsars of Q and M class. 
The procedure started with a
 selection of as many Q and M-type profiles as possible,
followed by a visual estimate of component number and location.
The objects have been selected from the works of R93, G93, 
Hankins and Rankin (2010), and 
the other sources itemised in the caption to Table \ref{tabobs}.
\nct{ran93, gks93, hr10}
Their profiles were grouped
into three classes of quality (mostly determined by the easiness
 to discern components), with the worst group rejected.
Profiles of the remaining $30$ objects, listed in Table \ref{tabobs}, were viewed
at a large (clear) scale, with four vertical lines overplotted at the suspected
locations of components' peaks. In five-component profiles the central one
has been ignored. The coordinates of the vertical lines
were used to calculate $\win$, $\wout$ and $R_W=\win/\wout$ given 
in Table \ref{tabobs}. In some cases (like B1821$+$05) it was helpful
to refer to two frequencies to resolve doubts about the existence
or location of a specific component.

The profiles have purposedly been not decomposed 
by fitting analytical functions, mainly because such functions
are unknown, and different components are apparently described
by different analytical shapes (cf.~fig.~1 in Kramer et al.~1994
with the outer components of J0631$+$1036 in fig.~2 of Weltevrede et al.~2010).
\nct{kwj94, waa10} 
Intrinsic intramagnetospheric
effects often appear to make the components asymmetric, and the drifting
phenomenon can possibly make them roughly triangular or trapezoidal.
A fitting of the Gauss curves can therefore give false results, 
biased by the use of the incorrect function in the decomposition.
In the case of blended asymmetric components, 
the unguided 
Gaussian fitting is incapable to provide precise estimate 
of the components'
positions or even of their number (the latter must be decided by eye
also for isolated asymmetric components, see the fit of the von Misses
functions in fig.~2 of Weltevrede \& Johnston 2008). 
\nct{wj08}
For these reasons we have identified the components
visually. The rather large width of bins used in our
observed $R_W$-histogram ($\Delta R_W=0.05$) makes the analysis
less sensitive to errors. Moreover, the difference in the predictions of the
models will be shown to be so large, 
that even the
approximate estimate of the observed $R_W$ histogram is useful.

 Our identification of the morphological type (Q or M) is based solely
on the number of easily identifiable components, so it may be different 
from that of R93, who also considered spectral and circular polarisation
properties. Out of the 17 pulsars common for our Tab.~\ref{tabobs}
and R93, only 8 have been designated a definite type 
in R93.  Three of those (B0621$-$04, B1845$-$01 and B1918$+$19)
have a different type assigned (respectively M, cT, and cT in R93). 
If all the three objects are rejected from the analysis, the observed 
$R_W$ distribution retains its boxy shape with little impact 
on our conclusions.

\section{Conal model versus data}
\label{concom}

\begin{figure}
\includegraphics[width=0.48\textwidth]{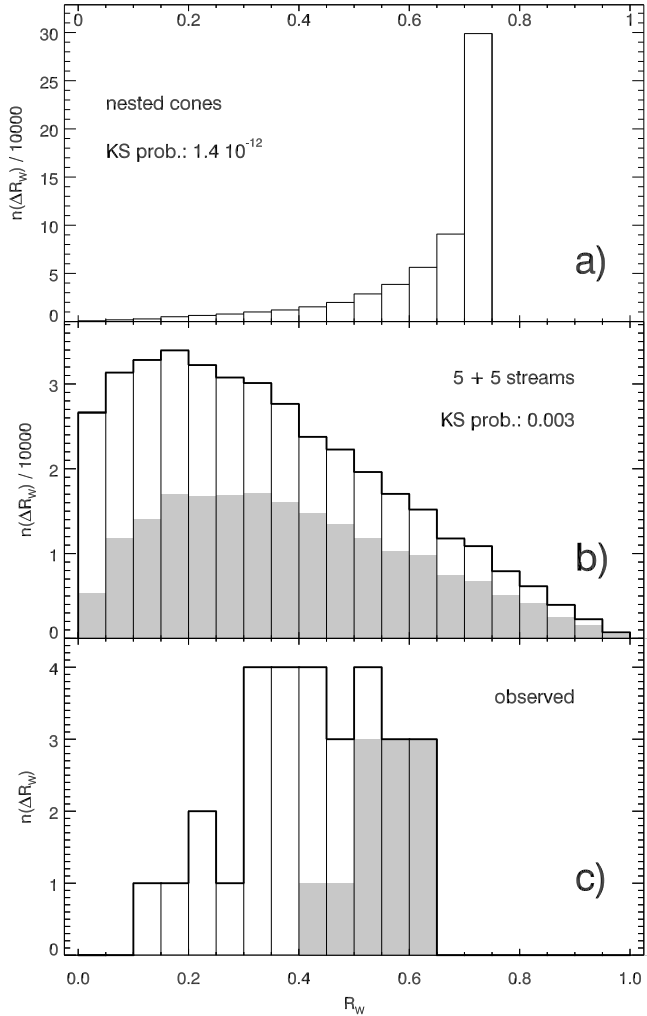}
\caption{Comparison of the simulated $R_W$ distributions 
(conal model in panel a; fan beam model in b)
with the one observed for Q and M type pulsars (panel c).
The numbers in a and b give the KS probablity of consistency
with the observed distribution. Note the low likelihood
of the conal model. Grey parts of the histograms in b and c 
denote the M-type profiles. An unknown fraction of these is also included in
the histogram of panel a, which shows the sum of the Q and M profiles. 
}
\label{com}
\end{figure}

The observed distribution is presented in 
Fig.~\ref{com}c, whereas the one simulated for
 the nested-cone model -- in Fig.~\ref{com}a. 
The distributions are completely different. The conal model
distribution is dominated by the value $\rwcon=R_\rho$, which corresponds to
the beam size ratio. The observed $n(\rwobs)$ peaks at the value of $\rwobs
\sim 0.375$ which should have been nearly absent in the data.
 The Kolmogorov-Smirnov (KS) test (Press et al.~1992) \nct{ptv92} 
excludes the common origin 
of the distributions, giving it a probablity of $10^{-12}$.

The total distribution of the conal model is marginally consistent 
(common origin prob.~of 0.04) 
with the M-type part of the distribution alone (grey part of the histogram 
in Fig.~\ref{com}c). However, we find no convincing reason to argue that
the nested cone structure is only responsible for the observed M-type
profiles, whereas the Q profiles have different origin.
The ratio of Q to M pulsar numbers is difficult to estimate in the conal
model, because the detectability of the core depends on the central beam
parameters and telescope sensitivity. Therefore we focus on the total
distributions.

Since the theoretical distribution is dominated by the peak at $R_\rho$,
the observed $n(\rwobs)$ could only be explained by the nested-cone beams 
if most of them has the size ratio $R_\rho \sim 0.3 - 0.65$ with
most common values within $(0.3,0.45)$.
This is not consistent with the findings described in the introduction
(Tab.~\ref{tab1}).
Even in the case of MD, who tentatively identified additional 
large cone with $\rho=1.3\rout$, their data are dominated by the ratio
$R_\rho = 0.8$, inconsistent with the observed statistics of $\rwobs$.
Furthermore, we are not aware of any magnetospheric arguments, 
which could support such a large variety of $R_\rho$, as implied by
Fig.~\ref{com}c. 

\subsection{Selection effects in the conal model}
\label{selcon}

\begin{figure}
\includegraphics[width=0.48\textwidth]{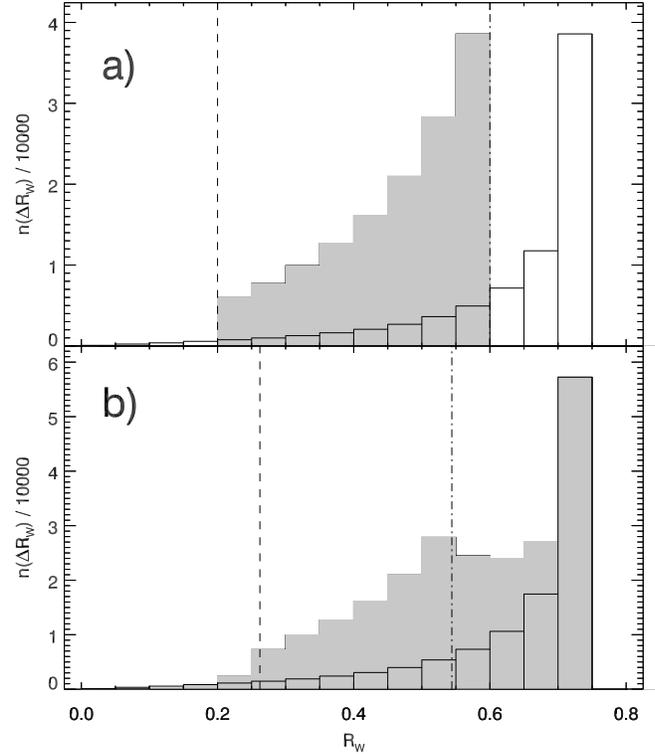}
\caption{The influence of selection effects caused by component blending
on the conal $R_W$ distribution. The grey histograms are affected 
by the limited resolving capabilities: {\bf a)} $\dres = 0.2\wout$;
{\bf b)} $\dres = 2.1^\circ$. The width of the affected histograms
(or their humps) can be estimated from eqs.~\mref{con1} and \mref{con2}
(dashed and dot-dashed vertical line, respectively).
The unaffected histogram from Fig.~\ref{com}a is shown as a reference.
}
\label{cutobs}
\end{figure}

Individual observed components may be difficult to distinguish 
because of their considerable width, or a large noise
 (instrumental or intrinsic, ie.~related to a large dynamic range
 of dissimilar single pulses).
It is therefore reasonable 
to introduce a threshold for their minimum resolvable proximity 
$\dres$.
This can be defined either as a fraction $\epsilon_W$ of the total pulse
width $\wout$ or as an absolute value measured in degrees.
In the first case the $R_W$ histogram 
becomes narrower and less spiky,
as shown in Fig.~\ref{cutobs}a.
The part of the histogram left of the vertical dashed line
becomes empty because $\win<\epsilon_W \wout$ (ie.~$R_W < \epsilon_W$)
for the peripheral cut through the
beam. Provided $\epsilon_W > (1-R_\rho)/2=0.125$, the part on the right-hand
 side of the dot-dashed line also vanishes 
because the distance between adjacent inner and outer components 
($(\wout-\win)/2$) is smaller than the resolution $\epsilon_W \wout$
for the central cuts through the beam. The result of Fig.~\ref{cutobs}a 
does not depend on
$\alpha$ since the resolution is scaled just as the profile width.
For $\epsilon_W = 1/3$ the histogram contracts to the zero width 
(the dashed and dot-dashed lines meet 
at $R_W = 1/3$) and no components can be resolved in a profile.

Assuming that we are incapable to resolve components at a distance
smaller than $\epsilon_W \wout$, the conal distribution
can be made consistent with observations for $\epsilon_W \approx 0.22$
(common origin KS probablity: $0.58$; the probability stays above $10^{-2}$
within $0.18<\epsilon_W<0.25$).
The observed distances between adjacent components
 ($\Delta\phi_{\rm adj}$)
indeed have a distribution (not shown) which decreases steeply
below $\Delta\phi_{\rm adj}/\wout \approx 0.2$. It may therefore be possible
that the observed distribution of $R_W$ (Fig.~\ref{com}c) is heavily distorted
by our limited capability to resolve intrinsically overlapping components.
Such profiles, however, (with the outer components merged with their inner
neighbours), should either have two well-separated components (D class),
or three components with the central one well-separated (by 
$0.375\wout$) from the inner conals (class T). These should be numerous, because
they represent the right hand-side peak in the conal histogram.
This would imply that conal components in several profiles of D and T class 
should consist of two (inner and outer) blended components.

In the case of the fixed ($\wout$-independent) resolving capability
the shape of 
the $R_W$ histogram (obtained for all combinations of isotropically
distributed $\alpha$ and $\zeta$)
depends
on the value of $\dres$ 
(equal to $2.1^\circ$ in Fig.~\ref{cutobs}b).
 The reason can be readily seen by decomposing the total $R_W$ histogram 
into subhistograms which correspond to a fixed value of $\alpha$
(or a narrow interval of $\alpha$) and the isotropic $\zeta$.
For a small $\alpha$, profiles are wider because they are viewed 
at small angles with respect to the rotation axis, hence, 
 the resolving limitations 
apply only for narrow profiles observed at a larger 
$\zeta$ and $\alpha$. 
The resulting $R_W$ histogram is then composed of the unaffected
part (with the sharp peak at $R_W = \rat$, corresponding to the circumpolar
viewing at a small $\zeta$) 
and a range of narrower sub-histograms corresponding to 
larger viewing angles (affected by $\dres$). 
The strongest (most numerous), and narrowest contribution 
comes from the equatorial viewing (cases with $\zeta\sim\alpha\sim90^\circ$)
and is visible in Fig.~\ref{cutobs}b as the protruding part 
between the dashed and dot-dashed lines.

Locations of the resulting bumps in the histogram can easily be determined
analytically.
The dashed line constrains the region where
$\win \ge \dres$, which corresponds to
\begin{equation}
\beta^2 \le \rin^2 - \frac{\dres^2}{4}, \ \ \ \ \ \ 
R_W^2 \ge \frac{\dres^2}{4(\rout^2 - \rin^2) + \dres^2}.
\label{con1}
\end{equation}
The dot-dashed line sets the upper limit for the region wherein 
$(\wout-\win)/2 \ge \dres$, ie.:
\begin{eqnarray}
\beta^2 & \ge & \rin^2 - \frac{(\dres^2+\rin^2-\rout^2)^2}{4\dres^2},\\
R_W & \le & \left|\frac{\dres^2+\rin^2-\rout^2}{\dres^2+\rout^2 -
\rin^2}\right|.
\label{con2}
\end{eqnarray}
In the case of large $\dres$ and $\alpha\simeq90^\circ$,
the limiting conditions \mref{con1} and \mref{con2} may exclude the entire
parameter space (all profiles unresolved/rejected, $\Delta\phi_{\rm
adj} < \dres$ for any $\beta$). This happens
 when $\dres^2 = (\rout^2-\rin^2)/2$ and $R_W = 1/3$. The histogram
then has a single bump (or break) at $R_W = 1/3$,
and the two vertical lines in Fig.~\ref{cutobs} coincide.
The actual look of a distribution affected by the limited resolution
depends on the relative value of $\dres$ as compared to the scale of the
beam. The width of the humps (or of the histogram itself) 
will change whenever 
parameters such as the $\nu$-dependent $r$, the rotation period, 
or the lateral boundaries of the cones are changed.
Tests performed for different values of the fixed $\dres$, 
have shown that the undistorted
part of the histogram (with the peak at $R_W=0.75$) 
usually contributes considerably to the overall shape,
and the probability of consistency 
can hardly exceed $6\times 10^{-4}$ (at $\dres = 2.5^\circ$ and $P=1$ s).

The conal model is then found a poor representation 
of data, unless a properly-tuned selection effect is called for:
it should be impossible to resolve components separated by less than
$0.2\wout$, regardless of the profile width $\wout$.
This is possible if the intrinsic width of components, responsible for the
blending, corresponds to a fixed fraction of polar tube.
However, for the best-fit value of $\epsilon_W = 0.22$, the number of profiles
with unresolved peripheric components (that would be classified as
the apparent double D or triple T profiles) is four times larger
than the total number of known (resolved) Q and M profiles. 
That would imply that
for some $\sim 200$ pulsars of D and T type, the outer components
are composed of the unresolved pairs. 


\section{Peak separation ratio in the stream model}
\label{Peak separation stream}

\begin{figure}
\includegraphics[width=0.48\textwidth]{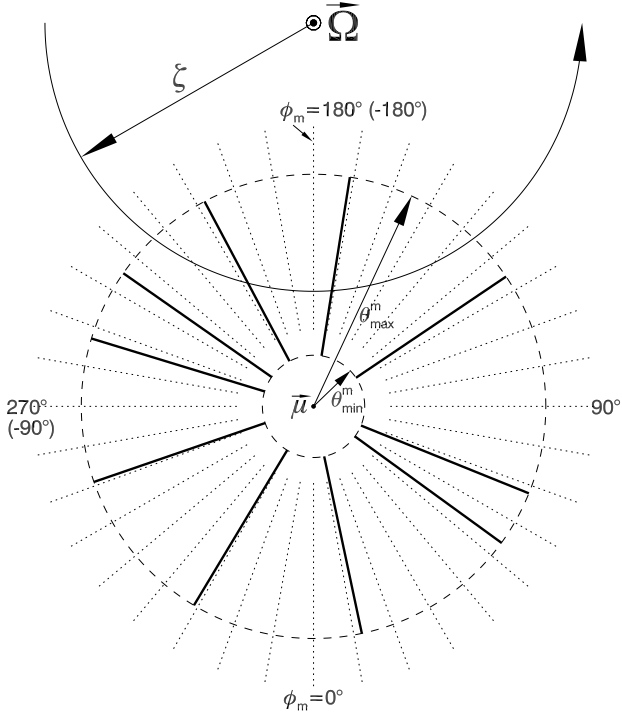}
\caption{A view of a simple fan beam down the dipole 
axis $\vec \mu$.
The thick radial sections emit radio waves within the range of 
magnetic angles $\thm \in(\thmin, \thmax)$. 
The observer's line of sight revolves
around the star rotation axis $\vec \Omega$, along the solid line path
(arc with the arrow). The distance between $\vec \Omega$ and $\vec \mu$
corresponds to the dipole tilt $\alpha$. 
Poleward viewing is presented
($\sin\zeta < \sin\alpha$) of a pulsar with small $\alpha$. In the 
case shown, a two-component profile of D class would be recorded. 
Note the definition of the magnetic
azimuth $\phm$. The dotted projections of magnetic field lines
are separated by $10^\circ$ in $\phm$. 
} 
\label{azimdef}
\end{figure}

To learn the shape of the $\rwstr$ distribution for the stream model
of pulsar beam (DRD10; Dyks \& Rudak 2012; Wang et al.~2014)
\nct{drd10, dr12, wpz14}
we consider the simple geometry of emission limited to separate magnetic
azimuths $\phm$. Nearly all observed radio profiles have at most five
components. Therefore,
for each beam in the sample, ten radio-emitting streams, in a `5+5' fashion,
 is assigned to the polar region: five of them 
in the upper (poleward) half of the polar tube,
and the other five in the lower (equatorward) part (see Fig.~\ref{azimdef}).
The magnetic azimuths $\phmi$ were selected randomly. 
For each a\-zi\-muth, a uniform radio emission was assumed within a limited
range of angles
$\thm$ measured from the dipole axis: $\thmin<\thm<\thmax$. 
As before, the values of $\alpha$ and $\zeta$ have been sampled isotropically.
The analysis was limited to Q and M profiles, ie.~we discarded
all the cases with less than four intersections between the azimuth of the 
emitted beam and the sightline path. Cases with more than five crossings,
which rarely appear for the extremely aligned geometry 
($\alpha\sim\zeta\lap \thmax$), have also been ignored.
The component separations $\win$ and $\wout$ have been calculated 
for the inner and outer pair of the crossing points, respectively
(in the cases with five intersections, the central one, corresponding to
the `core' component, was ignored).
The widths $W_i$ have been calculated using the strict spherical
trigonometric formalism, as described in DRD (see eq.~19
therein). \nct{drd10}
Some technical 
complications are also discussed in Appendix D of the present paper.

The $R_W$ distribution calculated for the stream model is shown in
Fig.~\ref{com}b. It is different from the observed one (KS probability of
consistency: $0.002$), albeit it is nine orders of magnitude more probable 
than the raw conal distribution (by `raw' we mean the distributions 
unaffected by the limited resolving capability). 
As in the observed case, the fraction 
 of the M type objects (grey part) increases with $R_W$.
This is because the extra space needed for the central component makes 
the leading side
components (outer and inner) more distant from the trailing pair. 
 Accordingly, $R_W$ is closer to unity. 
The increased fraction of M-type profiles at large $R_W$
 is also expected for the conal model, because detection of the core
requires a more central traverse through the beam.
It needs to be emphasized, however, that the number of radio-emitting 
streams that exist in magnetospheres of different pulsars likely varies between  
$0$ and $\sim\negthinspace\negthinspace 5$ (per one, poleward or equatorward, magnetic hemisphere).
The cases with the small numbers of streams (1, 2, 3) 
are likely responsible for
majority of the single, double and triple profiles.
Therefore, the observed ratio of M and Q profiles, is also affected
by the ratio of objects that actually have 4 or 5 streams, 
and not only by the statistics of the traverse through the beam with
5-streams.\footnote{The multiparameter modelling of the relative numbers 
of different
profiles (of S, D, T, Q, and M class), 
with the allowance for different numbers of streams in different objects,
is a complicated subject which will be discussed elsewhere (Frankowski et
al.~2016, in preparation; 
see also Karastergiou \& Johnston 2007).\nct{kj07}}

\subsection{Selection effects in the stream model}

\begin{figure}
\includegraphics[width=0.48\textwidth]{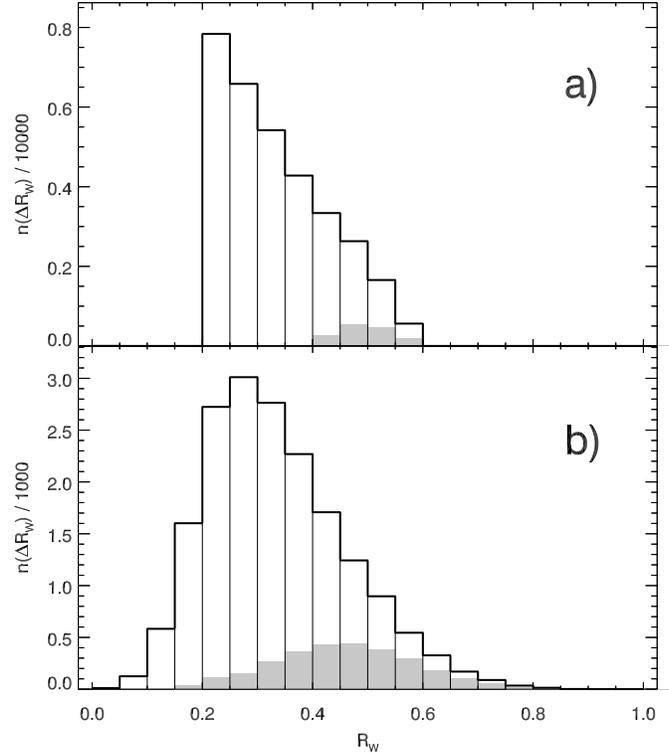}
\caption{The influence of component blending on the $R_W$ distribution in
the fan beam model. In {\bf a)} the minimum distance of resolvable
components is $\dres = 0.2\wout$. In {\bf b)} $\dres=1.5^\circ$. 
Note that the limited resolution of blended components
strongly increases the fraction of Q type profiles (white part of the
histograms). 
}
\label{cutobs7}
\end{figure}

As before, two types of selection effects have been applied to the 
stream model: the incapability to resolve
components located closer than $\epsilon_W\wout$ and the 
$W$-independent resolution limit of $\dres$ (see Fig.~\ref{cutobs7}a and b, 
respectively). The effect of these on the $R_W$ distribution was
similar to the one described for the conal model: in the case of the resolution
proportional to the width 
(fixed $\epsilon_W$, Fig.~\ref{cutobs7}a) the distribution becomes narrower,
and subdistributions of $R_W$ calculated for different $\alpha$ are the same
($\alpha$-independent). At $\epsilon_W=0.11$ the probability of consistency
with data reaches $0.04$.

For a fixed $\dres$ the selection effects do not
operate at small $\zeta$ so the resulting distribution consists
of several contributions of different widths (Fig.~\ref{cutobs7}b).
The fixed-$\dres$ distribution is narrower, however, 
probably on the account of its leftward skewness,
it is not very consistent with the observed one (Fig.~\ref{com}c) 
at any value of $\dres$ (maximum probability of consistency in the KS test: 
$0.016$ at $\dres=1.1^\circ$). 

\subsection{Influence of emission region geometry}

Contrary to the conal model (see Sect.~\ref{intro}), 
the geometry of the stream-shaped
region of radio emission is not even weakly constrained, either by theory
or observations.
Therefore, we have probed parts of the available parameter space 
by varying the following parameters:
1) the minimum angular distance of the emitting streams
from the dipole axis ($\thmin$, which has so far been set to zero);
2) a minimum azimuthal distance of the streams in the magnetic azimuth
$d\phm$;
3) an interval $\Delta\phm$ of magnetic azimuths that are
available for positioning the streams, ie.~
the original choice of the equatorward interval $-90^\circ<\phm<90^\circ$, and 
the poleward one $180^\circ-90^\circ<\phm<180^\circ+90^\circ$,
has been replaced with two narrower intervals which do not extend that far
from the main meridian: $-\Delta\phm<\phm<\Delta\phm$
and $180^\circ-\Delta\phm<\phm<180^\circ+\Delta\phm$, with
$0<\Delta\phm<90^\circ$.

 The value of $\thmin$ was varied in the full range between $0$
and $\thmax$. 
With the increase of $\thmin$ the original distribution 
(shown in Fig.~\ref{com}b) transforms into one which peaks near $R_W=0$
and decreases monotonically at larger $R_W$. The consistency with
observations stays at the level of a few$\times 10^{-3}$, until 
$\thmin$ reaches 75\% of $\thmax$. For a larger $\thmin$ the initially 
spoke-like shaped pattern of elongated streams 
starts to resemble 
`patchy cones', 
and the probability of consistency with data 
drops down to $2\times 10^{-4}$ at $\thmin=0.9\thmax$.
 For all results included in this paper, 
the value of $\thmax$ was set equal to 
the conal value of $\rout$, although we have also tried the low multiplicities
$i\rout$, with $i$ between $1$ and $5$.
Since $\thmax$ just rescales the beam,
the result does not depend on $\thmax$ as long as $\thmax \ll 1$ rad 
and the selection effects are neglected.

The limit for the minimum allowable azimuthal separation of streams
makes the agreement with data worse: the distribution tends towards
a narrow bump at $R_W=0.2$.

By keeping the streams closer to the main meridian 
($\Delta\phm=0.75 \times 90^\circ$),
the probability of consistency with the observations 
can be increased to a considerable value of $0.23$.
The distribution has the shape of a nearly symmetric, wide bump
with a peak at $R_W=0.4$. This is the simplest (single-parameter-based)
way to put the stream model into consistence with data at a considerable
 probability level. 

\subsection{Emission geometry plus selection effects}

\begin{figure}
\includegraphics[width=0.48\textwidth]{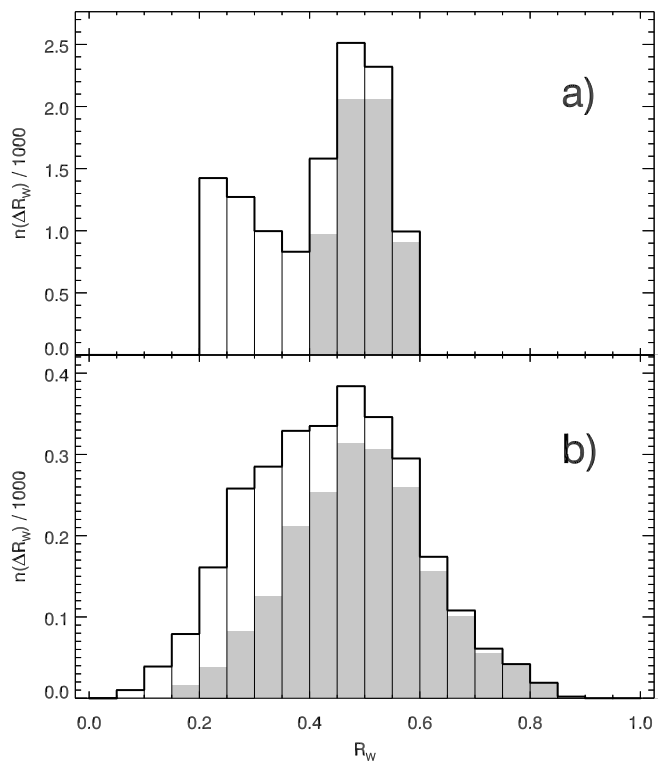}
\caption{Simultaenous effects of component blending 
and azimuthal confinement of streams. In both panels $\Delta\phm =
0.4\times90^\circ$, ie.~the streams are limited 
to two zones centered at the main meridian:
$\phm\in(-36^\circ, 36^\circ)$ and $(144^\circ, 216^\circ)$.
{\bf a)} $\dres = 0.2\wout$, {\bf b)} $\dres = 1.5^\circ$.
Note that such circum-meridional confinement of streams
increases the fraction of M type profiles (grey part of the histograms).
}
\label{cutobs11}
\end{figure}

An excellent agreement of the stream model with the data can easily 
be achieved
when both the geometric parameters and selection effects are simultaneously 
taken into account. Fig.~\ref{cutobs11}a presents a case with 
streams located closer to the main meridian
$\Delta\phm = 0.4\times 90^\circ=36^\circ$ and 
the blending unresolved below $0.2\wout$. Note the
change of the Q-to-M profile fraction (cf.~Fig.~\ref{cutobs7})
and the consistency of the total $R_W$ distribution with data
(KS probability of common origin: $0.77$).
Fig.~\ref{cutobs11}b presents a similar result 
($\Delta\phm = 0.4\times
90^\circ$) for a fixed longitude resolution $\dres=1.5^\circ$.
The probability of consistency with Fig.~\ref{com}c 
has increased up to $0.998$.
Note the peaks of the Q and M-type subdistributions in Fig.~\ref{cutobs11}a
are narrow and misaligned which produces a double-peaked  histogram.
In Fig.~\ref{cutobs11}b the subparts are misaligned but wider, hence
the shoulder at $R_W=0.3$.

\section{Conclusions}

We find that the markedly different types of radio emission region
(conal versus fan beam) imply dramatically different distributions
of peak separation ratio for profiles with $4$ and $5$ components.
In the conal model, the ratio $R_W$ is vastly dominated by values
close to the size ratio $\rat$ of the two cones. This has been shown to be
around $0.75$ in several published analyses. 
In the case of the stream model, a much wider distribution with
a peak at $R_W\approx0.2$ is predicted. 
The observed distribution is a broad bump centered at $R_W\approx0.4$.

When these simplest, raw predictions are compared to the 
observed distribution, the stream-based model is many orders of magnitude
more probable than the conal one.
However, this large difference between the predictions is mostly lost,
when selection effects and geometric parameters are allowed to enter
or change. The conal model can be made consistent with data
by invoking our incapability to resolve components located closer than
$0.22\wout$. Such resolution limit scales proportionally to
the profile width, which could only be interpeted as an intrinsic
effect of thick-walled emission rings occupying a fixed fraction of
polar tube. Such interpretation requires that a four times larger number
of pulsars (than the summed number of known Q and M profiles)
should be hiding the merged pairs of components from our view.
Such unresolved Q and M profiles should be observed as double
and triple ones.

The stream model (fan-beam model) can be made consistent with data
when the streams are positioned closer to the main meridian.
A good agreement has been achieved for a zone consisting of two parts
 (poleward and equatorward) centered at the main  meridian, each
of full width $\Delta\phm=72^\circ$.
When the selection caused by the component blending is also 
taken into account,
a perfect agreement of the stream model with data 
can be reached with no difficulty.

The one-dimensional distribution of the peak separation 
ratio is a new and interesting probing tool, which is free from (or mostly
insensitive to) several
unknown parameters, such as emission altitude of emission region or dipole
tilt.
Nevertheless, the method depends on the observational capability to resolve
blended components, which makes it less definite than the raw
predictions of Fig.~\ref{com}. Still, our results
 corroborate the success of the fan beam model in reproducing the properties
of pulsar profiles. On the contrary, the conal model is allowed to persist
 only under specially tuned circumstances.

 The poor performance of the nested cone model under the $R_W$ test,
deserves a discussion in view of its previous success in reproducing
the pulse-width distributions. The conal methods, as well as the
recent method of Rookyard et al.~(2015), 
assume that the lateral width of the radio beam decreases with the
increasing impact
angle $\beta$, which is not the case for the fan beam model.
 The method of Rookyard et al.~allows for a beam which is to a large degree 
arbitrary (eg.~patchy), however, since they associate the leading- or trailing-side
outskirts of a pulse profile with the \emph{circular} crossection 
of the polar tube, their method shares important qualitative properties with the
conal model. All those methods  imply a nonisotropic 
distribution of the dipole tilt, with moderately small  values of $\alpha$ 
preferred (Tauris \& Manchester 1998; Rookyard et
al.~2015; and the references in Tab.~\ref{tab1}).
\nct{tm98,rwj15}
The 
anisotropic $\alpha$ distribution is a valid possibility,
since the radio emissivity may depend on $\alpha$,
just as expected from the accelerating electric field 
(Arons 1983; Harding \& Muslimov 1998).  \nct{a83,hm98}

Our method assumes the isotropic $\alpha$ distribution, 
however, 
this assumption may
not be biasing our results, because the $R_W$ method is 
essentially\footnote{Detailed shape of the $R_W$ distribution depends 
on $\alpha$ when the components' resolvability is 
limited by the fixed $\Delta\phi_{\rm
res}$ (Sec.~\ref{selcon}). The bump at a moderate $R_W$ is then present (Fig.~\ref{cutobs}b),
which corresponds to the cases with large $\alpha$ and $\zeta$. Therefore,
when the $\alpha$ distribution is assumed to concentrate at small values,
where the selection effects cease to operate, the bump in the 
$R_W$ distribution becomes less pronounced. The small-$\alpha$
preference then makes the histogram more discordant with the observations
than in the case of the isotropic $\alpha$ distribution.}
 independent 
of $\alpha$. It is therefore possible that the small-$\alpha$ preference,
as found by the other methods, results from the incorrect (circular) shape
 of the beam's outer boundary.
Qualitatively, this explanation would work in the right direction, for the
following reason. 
Profiles produced by fan-shaped beams usually widen with the increasing
impact angle $\beta$. This is the case when a single outflowing stream
extends laterally over an interval of the magnetic azimuth, 
or when there is more than
one stream. For a central passage of the sightline ($\beta = 0$)
the width of a profile formally vanishes ($W=0$).
Therefore, to reproduce a given observed pulse width $W$, a fan beam 
must be traversed at an appreciably large $\beta$. This, 
through the observationally-fixed slope of a polarisation angle curve
($S=\sin\alpha/\sin\beta$),
implies larger $\alpha$ than for the conal beam.
The conal beam model, on the other hand, may tend to underestimate 
the values of $\alpha$ and $\beta$, in its effort 
to fit the observed widths of profiles  (or core components) 
through the sightline traversing too close to the dipole axis.
Unless we allow for the strong selection effects, the $R_W$ method
is only sensitive to parameters that determine the beam shape (not the
scale). Therefore, we suggest the problems of
the traditional conal model are inherent to the model itself, 
rather than to our method. The problems are probably caused by an incorrect, 
or at least not universal beam shape.  

A population of the nested-cone beams with a fixed $\rat$,
stands out as a narrow spike at $R_W=\rat$ in the $R_W$ distribution.
The latter is observed to have the boxy shape with the $R_W$
in the range between $\sim\negthinspace0.3$ and $0.65$.
Therefore, the nested-cone beam can be made consistent with 
Fig.~\ref{com}c, if a similar range of $\rat$ is assumed to exist 
in the real population of beams (with different values of $\rat$ 
being comparably numerous). However, this would imply an average $\rat$
 of $\sim\negthinspace0.5$, inconsistent with the numbers cited 
in Tab.~\ref{tab1}.

The present study has been inspired by the very symmetric
four-component profile of J0631$+$1036 (Zepka et al.~1996; 
Teixeira et al.~2016; 
Weltevrede et al.~2010). \nct{zcw96, trw16, waa10}
The inner components of its profile
are located very close to each other, implying
$R_W=0.33$. As can be seen in Fig.~\ref{com}c, we have identified
three more objects with $R_W$ in the range between
$0.3$ and $0.35$. This makes up for $13$\% all objects in the observed $R_W$
histogram. According to the conal $R_W$ distribution
(Fig.~\ref{com}a) only $1.6$\% of all Q and M objects
should fall within $0.3<R_W<0.35$. As noted in Teixeira et al.~(2016), 
this suggests that the stream-based geometry (a system of fan beams) 
is responsible for the unusually symmetric profile of J0631$+$1036.
This idea is strongly supported by the presence of deep minimum
at the center of the profile, (between the inner components), where
the radio flux drops nearly to zero. This is difficult to understand
within the conal model, because for $R_W=0.33$ the nested cones 
with $\rat=0.75$ imply the impact angle $\beta=0.95\rin$, 
ie.~at the center of the profile
the sightline stays very close to the peak emissivity of the
inner cone. Therefore, the nearly vanishing central flux 
seems to be incompatible with the standard conal geometry.

Overall, the results discussed above provide more support for the fan beam
geometry, and raise more problems for the conal model.
In view of the other arguments for the stream model (DRD10; Dyks \& Rudak
2012; Wang et al.~2014; Chen \& Wang 2014; Dyks \& Rudak 2015)
\nct{drd10, dr12, wpz14, cw14, dr15}
it may be worth to shift the current paradigm of
geometric studies of pulsars from the conal structure towards
the azimuthal arrangement of fan beams.

\section*{acknowledgements}
JD appreciates generosity of A.~Karastergiou, S.~Johnston,
 and J.~M.~Rankin in providing us with pulsar profile data.
We thank B.~Rudak for reading the manuscript. 
This work was supported by 
the National Science Centre grant DEC-2011/02/A/ST9/00256.
\bibliographystyle{mn2e}
\bibliography{listofrefs}

\appendix

\section{The conal $R_W$-distribution in the flat geometry}
\label{flatderiv}

When the beam is narrow ($\rout \ll 1$ rad) and the dipole tilt is large
($\alpha \gg \rout$) the $R_W$ distribution is well approximated by the
flat geometry of Fig.~\ref{idea}.
In that case the components are detected at the pulse longitudes:
\begin{equation}
\phin^2 = \rin^2 - \beta^2, \hskip10mm \phout^2 = \rout^2 - \beta^2,
\label{phases}
\end{equation}
which determine the widths $\win=2\phin$ and $\wout=2\phout$.
The values of $R_W=\win/\wout=\phin/\phout$ and the impact angle $\beta$ 
are then given by:
\begin{equation}
R_W^2 = \frac{\rin^2 - \beta^2}{\rout^2 - \beta^2},
\hskip10mm  \beta^2 =\frac{\rin^2(1-R_W^2\rat^{-2})}{1 - R_W^2}.
\label{erwu}
\end{equation}
The number of observers that record $R_W$ within some interval 
of $\Delta R_W$ is proportional to the interval of impact angle 
$\Delta \beta$ which corresponds to that 
$\Delta R_W$. Therefore,
\begin{equation}
n(R_W) \equiv \lim\limits_{\Delta R_W \rightarrow 0}n(\Delta R_W) \propto \left|\frac{d\beta}{d R_W}\right|,
\label{en}
\end{equation}
which gives eq.~\mref{nodrw}. 

\section{Dependence of the conal $R_W$-distribution on the dipole tilt $\alpha$}
\label{alphadep}

To simplify the calculation, we consider a central-cut case with 
$\zeta = \alpha$ (ie.~$\beta = 0$).
Pulse longitudes of the inner and outer pair of components
are then given by:
\begin{equation}
\cos\phi_i = \frac{\cos\rho_i - \cos^2\alpha}{\sin^2\alpha}=
1 + \frac{\cos\rho_i-1}{\sin^2\alpha},
\label{cosphii}
\end{equation}
where $i=$ `in' or `out',
When eq.~\mref{cosphii} is Taylor-expandend to the 2nd power of
$\phi_i$ and $\rho_i$, one obtains $\phi_i \approx \rho_i/\sin\alpha$, 
so that
$R_W = \win/\wout = \rin/\rout=\rat$ does not depend on $\alpha$.
This explains the stability of the $R_W$ distribution 
visible in Fig.~\ref{alphas}a.

To recognize the weak $\alpha$-dependence of $R_W$, the $\cos\phi_i$ on the
left-hand side of
eq.~\mref{cosphii} needs to be expanded 
to the order of $\phi_i^4$. This leads to the quadratic equation for $\phi_i^2$:
\begin{equation}
\frac{\phi_i^4}{12} - \phi_i^2 + 
\frac{2(1-\cos\rho_i)}{\sin^2\alpha} = 0
\label{quadraticforphii}
\end{equation}
which has the solution:
\begin{equation}
\phi_i = \left[6\left(1 - \sqrt{1 - x_i^2}\right)
\right]^{1/2},
\label{phisol}
\end{equation}
with $x_i^2=2(1-\cos\rho_i)/(3\sin^2\alpha)\approx\rho_i^2/(3\sin^2\alpha)$.
Accordingly:
\begin{equation}
R_W^2 \approx \frac{1-\sqrt{1 - \xin^2}}{1-\sqrt{1-\xout^2}}
\approx \frac{1-\sqrt{1-\xout^2\rat^2}}{1-\sqrt{1-\xout^2}}.
\label{ratsq}
\end{equation}
Expanding the square roots around $\xout^2 \ll 1$ (up to the terms
$\propto\xout^4$) one arrives at 
\begin{equation}
R_W^2 \approx \rat^2\left(1 - \frac{1 - \rat^2}
{1 + 12\sin^2\alpha/\rout^2}\right).
\label{unity}
\end{equation}
Neglecting the unity in the denominator, taking a square root
and Taylor-expanding it up to $\rout^2/\sin^2\alpha$ gives the approximate
eq.~\mref{secorder}.
 This result shows that the value of $R_W$ depends on $\alpha$
only for a nearly-aligned geometry ($\alpha \la \rout$).
Therefore, the $R_W$ distribution is insensitive to 
the assumed distribution of $\alpha$, unless the latter 
is extremely non-isotropic, eg.~with most objects having $\alpha$ 
of a few degrees.

\section{Pulse longitudes for 
emission from a fixed magnetic azimuth}
\label{longitudes}

To compute $\win$, $\wout$, and $R_W$ for the stream model (fan beam model) 
one needs a prescription for how to calculate
pulse longitudes $\phi$ that correspond to the points where the
selected (radio-emitting) magnetic azimuths $\phm$ are sampled by the line of
sight.
A simple way to do 
this is to use eq.~(19) of DRD10:
\begin{equation}
\cos\zeta = \cos(\pi - \phm)\sin\alpha\sin\thm + \cos\alpha\cos\thm
\label{thm}
\end{equation}
to calculate the polar angles $\thm$ between the dipole axis and 
the emission direction from the sampled points
(ie.~the points at which the tangentially-emitting streams 
are detectable by the line of sight).
Then the pulse longitudes $\phi$ for each component (ie.~for each crossing point 
with some magnetic 
azimuth $\phm$) can be found in the usual way: 
\begin{equation}
\cos\phi = (\cos\thm - \cos\alpha\cos\zeta)(\sin\alpha\sin\zeta)^{-1}.
\label{phi}
\end{equation}
This would have been the full procedure, had it not been for
a few, following technical complications.

First, the cosine theorem of eq.~\mref{thm} leads to the following quadratic
equation
for $\cos\thm$:
$A\cos^2\thm + B\cos\thm + C = 0$, where:
\begin{eqnarray}
A & = & \cos^2\alpha + \cos^2\phm\sin^2\alpha\label{thma}\label{bla}\\
B & = & -2\cos\alpha\cos\zeta\label{thmb}\\
C & = & \cos^2\zeta - \cos^2\phm\sin^2\alpha.
\label{thmc}
\end{eqnarray}
For the positive discriminant $\Delta$, ie.~for $\sin^2\zeta >
(1-\cos^2\phm)\sin^2\alpha$, there are
two real solutions for $\theta_{m}$ which are measured from the same magnetic pole.
Since the result depends on $\cos^2\phm$ (eqs.~\ref{bla}, \ref{thmc}) 
there is a fourfold degeneracy
associated with $\phm$, ie.~the cases with $\pm\phm$ or $\pi\pm\phm$
cannot be discerned. Eqs.~\mref{thm} and \mref{phi} are thus
blind to the
sign of the magnetic azimuth (ie.~to the location of the stream on the leading 
or trailing side of the magnetosphere) so the sign of $\phi$ must be manually 
set equal to the sign 
of the corresponding $\phm$ (with the latter understood in the range of
$\pm\pi$).
Moreover, the result is insensitive to the continuation 
of a given magnetic meridian
to the other side of the magnetic pole (the solution for $\phm$ 
is the same as for $\phm+\pi$, ie.~it does not depend on the sign of $\thm$).
For example, consider a stream that starts on the poleward side 
of the dipole axis and extends away and upwards, into the
rotationally-circumpolar regions of the magnetosphere, at some fixed magnetic
azimuth $\phm$.
When the line of sight has $\sin\zeta > \sin\alpha$, it is passing on the equatorward
side of the dipole axis, so the poleward stream should be missed
 near the dipole axis, with no
achievable solution for $\phi$. However, the `$\phm$ plus pi' denegeracy
will result in a real solution for the extention of the azimuth $\phm$
 to the other
(equatorward) side of the magnetic pole ($\phm+\pi$). A simple way to reject these false
solutions is to insert all the calculated $\thm$ into eq.~\mref{thm} 
to check if the implied values of $\zeta_{\rm test}$ are 
consistent with the original $\zeta$.

For large dipole inclinations, the larger of the two solutions for $\cos\thm$
(hence related to a small $\thm$) 
corresponds to the stream crossing at a `near' magnetic pole.
The other solution has a smaller value of $\cos\thm$, with $\thm$ 
always measured from the same (near) magnetic pole.
In this latter case one may have $\cos\thm \sim -1$ and 
$\thm \sim180^\circ$ which corresponds
 to the passage through the same magnetic azimuth
close to the other (far) magnetic pole.
Since the radio emission is assumed to be latitudinally-limited 
to a single\footnote{The inclusion of the far magnetic pole 
would just renormalise
all the distributions by a factor of two.
Since we ignore the question of interpulses, the second pole is neglected.} 
 circumpolar range of $\thmin<\thm<\thmax\ll 1$ rad,
the far solution for $\cos\thm$ (the one for which $\thm \sim \pi$) is usually rejected.
However, in the case of the nearly aligned geometry
$\alpha\sim\zeta\lap\thmax$, both the solutions 
for $\thm$ (the smaller and the larger one) 
can survive at a single magnetic pole, ie.~our line of sight
can sample each $\phm$ twice, while staying within the 
radio-emitting zone of $\thm$.

Although we have had few nearly aligned (very wide) profiles in our observed
sample (eg.~B1831$-$04, see Table \ref{tabobs}), the complications resulting from the nearly 
aligned geometry 
(small $\alpha$) have been carefully treated.
With five streams extending both into the poleward and equatorward
part of the magnetosphere, it is possible to record components with more
than five streams in the nearly aligned geometry.
Ten-component profiles are possible when the sightline crosses each stream
twice, as well as in the case when $\alpha \ll \thmax$
 and all ten streams are crossed once per period.
In such cases, the definition of the outer and inner pair components
becomes somewhat arbitrary, so we assume that the off-pulse region
encompasses the azimuth $\phm = \phi = 180^\circ$. The correct ordering of components
is then achieved when the leadingmost component
\emph{simultaneously} has the smallest pulse longitude and magnetic
azimuth 
(when both are defined in the range $(-180^\circ, 180^\circ)$).
In the case of $\alpha \approx 0$, and $\alpha<\zeta \lap \thmax$,
the component with the smallest value of $\phm$ is unique.
In general, however, e.g.~for $\alpha=\thmax/2$ and $\zeta < \thmax/2$,
there may exist pairs of components with the same value of $\phm$,
including two with the minimum $\phm$ but different $\phi$.  

Accordingly, our simulation of $R_W$, $\win$ and $\wout$ 
for the stream model has followed these steps:
1) Ten magnetic azimuths have been selected (five in the equatorward range
of $\phm \in  (-90^\circ, 90^\circ)$, the other five within $(90^\circ,
270^\circ)$. 2) The values of $\alpha$ and $\zeta$ have been isotropically 
selected; the cases with $|\beta| > \thmax$ have been ignored.
3) The two groups of solutions for $\thm$ from eq.~\mref{thm} 
have been calculated;
those which implied inconsistent $\zeta$ have been rejected.
Those which fell outside the $(\thmin, \thmax)$ interval
were also ignored.
4) The pulse longitudes have been calculated from \mref{phi}
with the sign of $\phi$ set the same as for the corresponding $\sin\phm$.
Note that taking the sign of sine instead of that of $\phm$
redefines (transforms) the interval of $\phm$ from the initial
$(-90^\circ,270^\circ)$ (useful for poleward/equatorward stream selection) 
into $(-180^\circ, 180^\circ)$ (useful for the component ordering with the
off-pulse at $\phm=\phi=\pi$). 
5) Both groups of solutions have been merged into a single one-dimensional
array, then sorted and indexed increasingly. 
This produces a one-dimensional vector
of solutions with the offpulse region 
encompassing $\phi=\phm=-180^\circ$. 
The leftmost (leadingmost) solution has the smallest
azimuths (both the magnetic and the rotational one) when both are 
defined in the range of $\pm\pi$.
6) Only the cases with $4$ or $5$ solutions 
(corresponding to Q and M profiles)
have been considered. 
The widths $\wout$ and $\win$ have been calculated as the phase differences
between appropriate components: $\wout=\phi_n -\phi_1$,
$\win=\phi_{n-1}-\phi_2$ where $n=4$ or $5$.
The central solution (core) was
only used to calculate separations between adjacent components
$\Delta\phi_{\rm adj}$, needed to consider the selection effects 
that result from 
the component blending. Otherwise the `core' has been ignored.

\end{document}